\documentclass[prl,twocolumn,letterpaper, superscriptaddress]{revtex4-1}
\usepackage{graphicx}% Include figure files
\usepackage{dcolumn}% Align table columns on decimal point
\usepackage{bm}% bold math
\usepackage{color}
\usepackage{tabularx}
\usepackage{array}
\usepackage{amsmath}
\usepackage{stmaryrd}  %\llbracket \rrbracket

\begin{document}

\title{Coherent spin pumping in a strongly coupled magnon-magnon hybrid system}

\author{Yi Li\thanks{The first author wants to thank someone.}}
\affiliation{Department of Physics, Oakland University, Rochester, MI 48309, USA}
\affiliation{Materials Science Division, Argonne National Laboratory, Argonne, IL 60439, USA}

\author{Wei Cao}
\affiliation{Materials Science and Engineering, Department of Applied Physics and Applied Mathematics, Columbia University, New York, New York 10027, USA}

\author{Vivek P. Amin}
\affiliation{Maryland Nanocenter, University of Maryland, College Park, MD 20742, USA}
\affiliation{Physical Measurement Laboratory, National Institute of Standards and Technology, Gaithersburg, Maryland 20899, USA}

\author{Zhizhi Zhang}
\affiliation{Materials Science Division, Argonne National Laboratory, Argonne, IL 60439, USA}
\affiliation{School of Optical and Electronic Information, Huazhong University of Science and Technology, Wuhan 430074, China}

\author{Jonathan Gibbons}
\affiliation{Materials Science Division, Argonne National Laboratory, Argonne, IL 60439, USA}

\author{Joseph Sklenar}
\affiliation{Department of Physics and Astronomy, Wayne State University, Detroit, MI 48202, USA}

\author{John Pearson}
\affiliation{Materials Science Division, Argonne National Laboratory, Argonne, IL 60439, USA}

\author{Paul M. Haney}
\affiliation{Physical Measurement Laboratory, National Institute of Standards and Technology, Gaithersburg, Maryland 20899, USA}

\author{Mark D. Stiles}
\affiliation{Physical Measurement Laboratory, National Institute of Standards and Technology, Gaithersburg, Maryland 20899, USA}

\author{William E. Bailey}
\email{web54@columbia.edu}
\affiliation{Materials Science and Engineering, Department of Applied Physics and Applied Mathematics, Columbia University, New York, New York 10027, USA}

\author{Valentine Novosad}
\affiliation{Materials Science Division, Argonne National Laboratory, Argonne, IL 60439, USA}

\author{Axel Hoffmann}
\email{Current address:
Department of Materials Science and Engineering, University of Illinois at Urbana-Champaign Urbana, IL 61801
Email: axelh@illinois.edu}
\affiliation{Materials Science Division, Argonne National Laboratory, Argonne, IL 60439, USA}

\author{Wei Zhang}
\email{weizhang@oakland.edu}
\affiliation{Department of Physics, Oakland University, Rochester, MI 48309, USA}
\affiliation{Materials Science Division, Argonne National Laboratory, Argonne, IL 60439, USA}

\date{\today}

\begin{abstract}

We experimentally identify coherent spin pumping in the magnon-magnon hybrid modes of yttrium iron garnet/permalloy (YIG/Py) bilayers. {By reducing the YIG and Py thicknesses, the strong interfacial exchange coupling leads to large avoided crossings between the uniform mode of Py and the spin wave modes of YIG enabling accurate determination of modification of the linewidths due to the dampinglike torque.} We identify additional linewidth suppression and enhancement for the in-phase and out-of-phase hybrid modes, respectively, which can be interpreted as concerted dampinglike torque from spin pumping. {Furthermore, varying the Py thickness shows that both the fieldlike and dampinglike couplings vary like $1/\sqrt{t_{Py}}$, verifying the prediction by the coupled Landau-Lifshitz equations.}
\end{abstract}

\maketitle

Coherent phenomena have recently become an emerging topic for information processing with their success in quantum computing \cite{DevoretScience2013,AwschalomScience2013}. In
spintronics, exchange-induced magnetic excitations, called spin
waves, or magnons \cite{DysonPR1956,ChumakNPhys2015}, are good candidates for coherent information processing because
information can be encoded by both the amplitude and the phase of spin
waves. For example, the
interference of coherent spin waves can be engineered for spin wave
logic operations
\cite{KhitunIEEE2008,SchneiderAPL2008,KruglyakJPD2010}; the coherent
interaction of spin-torque oscillators leads to mutual synchronization
\cite{KakaNature2005,MancoffNature2005,locatelliSREP2015,LiPRL2017,LebrunNComm2017,AwadNPhys2017},
which can be applied in artificial neural networks
\cite{VodenicarevicSREP2017,RomeraNature2018}; and the coherent coupling
between magnons and microwave cavities
\cite{HueblPRL2013,TabuchiPRL2014,ZhangPRL2014,GoryachevPRApplied2014,BaiPRL2015,LiPRL2019_magnon,HouPRL2019,McKenziePRB2019}
opens up new opportunities for magnon-based quantum information
science \cite{TabuchiScience2015,LachanceScienceAdvan2017}.

Recently, strong coupling between two magnonic systems has enabled
excitations of forbidden spin wave modes \cite{KlinglerPRL2018,ChenPRL2018,QinSREP2018} and high group velocity of
propagating spin waves \cite{LiuNComm2018,AnPRApplied2019}. The coupling
is dominated by the exchange interaction at the interface of the
magnetic bilayers, providing a new pathway to coherently
transfer magnon excitations between two magnetic systems possessing
distinctive properties: from conductor to insulator, from uniform to
nonuniform mode and from high-damping to low-damping systems. However, the underlying
physical mechanisms of the coupling are still not fully
understood. First, what are the key parameters that dictate the
coupling efficiency and enable one to reach the strong-coupling regime? Second,
with the interfacial exchange coupling acting as a fieldlike torque,
is there a dampinglike torque associated with spin pumping
\cite{TserkovnyakPRL2002,HeinrichPRL2003,LenzPRB2004,TserkovnyakRMP2005}? {To resolve both questions, large separations of the two hybrid modes are required in order to quantitatively analyze the coupling mechanism.} The second question is also important for optimizing the coherence of spin wave transfer in hybrid systems.
Furthermore, the parasitic effect on the incoherent spin current from the conduction band is absent \cite{ZhangPRB2012,AminPRL2018,ChienPRB2019}
when using magnetic insulators such as yttrium iron garnet
(Y$_3$Fe$_5$O$_{12}$, YIG) \cite{MiaoPRL2013,HydePRB2014,AnPRApplied2019}, which facilitates the study of spin pumping coherency.

In this work, we study YIG/permalloy (Ni$_{80}$Fe$_{20}$, Py) bilayers. {By using much thinner YIG and Py films than studied in previous works \cite{KlinglerPRL2018,QinSREP2018}, we achieve an exchange-induced separation of the two hybrid modes much larger than their linewidths, allowing us to study the evolution of their linewidths in the strong coupling regime.}
We find a pronounced suppression of the total linewidth
for the in-phase hybrid modes and a linewidth enhancement for the
out-of-phase hybrid modes. The linewidths can be
understood from the Landau-Lifshitz-Gilbert (LLG) equation with
interfacial exchange coupling and mutual spin pumping, which provide the
fieldlike and dampinglike interlayer coupling torques, respectively.
{Furthermore, the thickness dependence of the two coupling strengths agrees with the modeling of coupled LLG equations with mutual spin pumping. The sign of the fieldlike torque also reconfirms that the YIG and Py are coupled antiferromagnetically \cite{KlinglerPRL2018}.}
Our results provide important insights for
improving the coupling strength and coherence in magnon-magnon hybrid
systems and pave the way for coherent information processing with
exchange coupled magnetic heterostructures.

%In contrast to the traditional computing schematics where information is encoded by the binary amplitude of charge current, the new schematic emphasize both the amplitude and the phase for information encoding, and the information carriers need to be an alternating excitation instead of electron charges.

\begin{figure}[htb]
 \centering
 \includegraphics[width=3.0in]{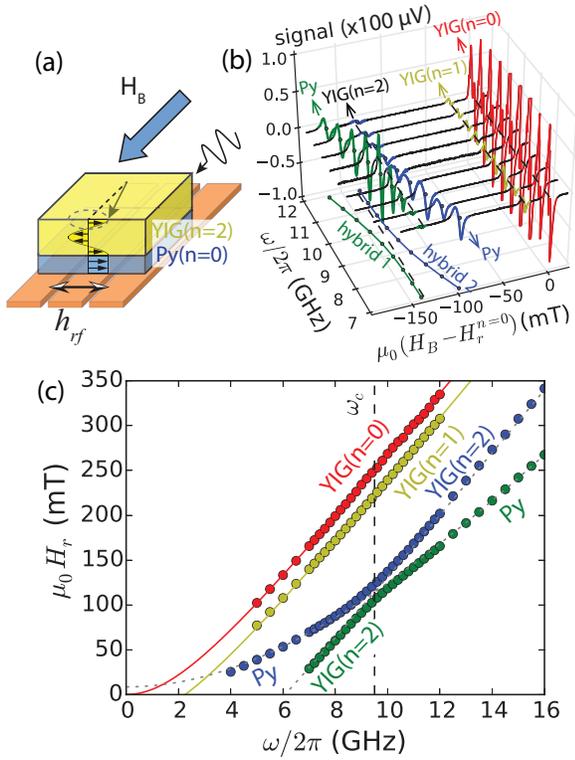}
 \caption{(a) Illustration of the {magnetization excitations} in the YIG/Py
   bilayers with a coplanar waveguide.
   (b) Lineshapes of the YIG(100~nm)/Py(9~nm) sample for the first
   three {resonance} modes of YIG and the uniform mode of Py. The field
   axis is shifted so that the resonance field of the YIG(n=0) mode is zero.
   (c) Unshifted evolution of the four modes in (b). Curves show the fits as uncoupled
   modes. The vertical dashed line denotes where the YIG($n=2$) and
   Py($n=0$) modes cross on the frequency axis at $\omega_c/2\pi=9.4$
   GHz. }
 \label{fig1}
\end{figure}

The samples consist of YIG(100~nm)/Py($t_\text{Py}$) bilayers where
$t_\text{Py}$ varies from 5~nm to 60~nm. YIG(100~nm) films were deposited by magnetron
sputtering from a YIG target onto Gd$_3$Ga$_5$O$_{12}$(111) substrates
and annealed in air {at 850$^\circ$C for 3 hrs} to reach low-damping characteristics
\cite{LiNanoscale2016}. Before the deposition of Py films on top of
YIG, the YIG surfaces were ion milled {\textit{in-situ}} for one minute in order to
enable good exchange coupling between Py and YIG \cite{JungfleischAPL2013}. For each Py
thickness, one additional reference Py film was deposited on a
Si/SiO$_2$ substrate during the same deposition.

The hybrid magnon dynamics were characterized by broad-band
ferromagnetic resonance with field modulation on a coplanar waveguide (Fig.~\ref{fig1}a). An
in-plane magnetic field $H_B$ saturates both the YIG and Py
magnetizations. Their Kittel modes, which describe spatially uniform
magnetization precession, are formulated as
$\omega^2/\gamma^2 = \mu_0^2 H_r(H_r+M_s)$, where $\omega$ is the mode
frequency, $\gamma/2\pi= (g_{eff}/2)\times 27.99$~GHz/T is the
gyromagnetic ratio, $H_r$ is the resonance field and $M_s$ is the
magnetization \cite{KittelPhysRev1948}.
For YIG, the spatially nonuniform perpendicular standing spin wave (PSSW) modes can be also measured.
An effective exchange field $H_{ex}$ will lower the resonance
field by $\mu_0H_{ex}(k)=(2A_{ex}/M_s)k^2$, where $A_{ex}$ is the
exchange stiffness, $k=n\pi/t$, $n$ labels the index of PSSW modes, and
$t$ is the film thickness \cite{KittelPhysRev1951}.

Fig.~\ref{fig1}(b) shows the line shapes of the resonance
fields for {the first three resonance modes of YIG ($n=0$, 1, 2)} and the Py
uniform mode ($n=0$) measured for $t_\text{Py}=9$~nm. For illustration,
the YIG ($n = 0$) resonance is shifted to zero field. An
avoided crossing is clearly observed when the Py uniform mode is degenerate
with the YIG ($n=2$) mode. This is due to the exchange
coupling at the YIG/Py interface
\cite{KlinglerPRL2018,ChenPRL2018,QinSREP2018} providing a fieldlike
coupling torque.
Both in-phase and out-of-phase YIG/Py hybrid modes are strongly excited because the energy of the Py
uniform mode is coherently transferred to the YIG PSSW modes
through the interface \cite{KlinglerPRL2018}.
The full-range frequency dependencies of the extracted resonance fields are plotted in
Fig.~\ref{fig1}(c).
To analyze the two hybrid modes, we analyze our results with two independent Lorentzians because it facilitates a transparent physical picture and the fit lineshapes agree well with our measurements.
The mode crossing happens at $\omega_c/2\pi=9.4$~GHz (black
dashed line), which corresponds to the minimal resonance separation of
the two hybrid modes. Fitting to the Kittel
equation, we extract $\mu_0M_s^\text{YIG}=0.21$~T,
$\mu_0M_s^\text{Py}=0.86$~T. From the exchange field offset as shown
in Fig.~\ref{fig1}(b), an exchange stiffness $A_{ex}=2.6$~pJ/m is
calculated for YIG, which is similar to previous reports
\cite{KlinglerJPD2014}.

The avoided crossing can be fitted to a phenomenological model of two
coupled harmonic oscillators, as previously shown in magnon polaritons
\cite{HueblPRL2013,TabuchiPRL2014,ZhangPRL2014,BaiPRL2015}:
\begin{equation}\label{eq01}
\mu_0H_c^\pm = \mu_0{H_r^\text{YIG}+H_r^\text{Py} \over 2} \pm \sqrt{ \left(\mu_0{H_r^\text{YIG}-H_r^\text{Py} \over 2}\right)^2 + g_c^2}
\end{equation}
where $H_r^{\text{YIG}(\text{Py})}$ is the resonance field of YIG
(Py), and $g_c$ is the interfacial exchange coupling
\textcolor{black}{strength}. $H_r^\text{YIG}$ and $H_r^\text{Py}$ are both functions of
frequency and are equal at $\omega_c$. Note that for in-plane biasing field the resonance field is nonlinear to the excitation frequency. This nonlinearity will be accounted for the analytical reproduction of Eq. \ref{eq01}. The fitting
yields $g_c=8.4$ mT for $t_\text{Py}=9$~nm.

\begin{figure}[htb]
 \centering
 \includegraphics[width=3.0in]{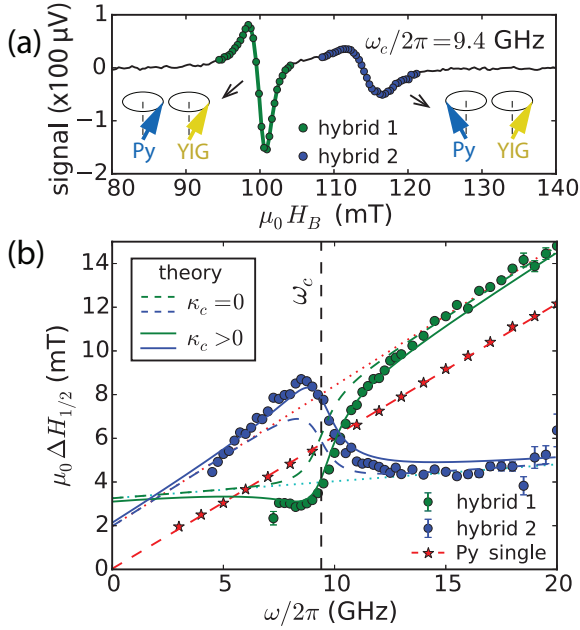}
 \caption{(a) The lineshape of the YIG(100~nm)/Py(7.5~nm) sample at
   $\omega_c/2\pi=9.4$~GHz, showing different linewidths between the two
   hybrid modes of YIG($n=2$) and Py($n=0$) resonances. (b) Linewidths
   of the two hybrid modes as a function of frequency. Dotted lines
   show the linear fit of the linewidths for the two uncoupled
   modes. Dashed curves show the theoretical values with
   $\kappa_c=0$. Solid curves show the fits with finite
   $\kappa_c$.}
 \label{fig2}
\end{figure}

%The interfacial exchange coupling between YIG and Py introduces not only a coherent coupling which shifts the resonance field of the hybrid modes, but also an incoherent coupling which shifts the lifetime of the two modes. The later corresponds to the spin pumping effect \cite{TserkovnyakPRL2002}, which brings a dampinglike torque with a $\pi/2$ phase offset to the fieldlike torque provided by interfacial exchange.

Next, we focus on the linewidths of the YIG-Py hybrid modes. Fig.~\ref{fig2}(a)
shows the line shape of the two hybrid modes for $t_\text{Py}=7.5$~nm
at $\omega_c/2\pi=9.4$ GHz (same value as for 9-nm Py). These two eigenmodes correspond to the in-phase and
out-of-phase magnetization precession of Py and YIG with the same
weight, so they should yield the same total intrinsic damping.
Nevertheless, a significant linewidth difference is
observed, with the extracted full-width-half-maximum linewidth
$\mu_0\Delta H_{1/2}$ varying from $3.5$~mT for the lower field
resonance to 8.0~mT for the higher field resonance. Fig.~\ref{fig2}(b)
shows the full-range evolution of linewidth. Compared with the dotted
lines which are the linear extrapolations of the YIG ($n=2$) and Py
linewidths, the linewidth of the higher-field hybrid mode (blue
circles) exceeds the Py linewidth and the linewidth of the lower-field
hybrid mode (green circles) reduces below the YIG linewidth when the
frequency is near $\omega_c$. This is the central result of the paper.
It suggests a coherent dampinglike
torque which acts along or against the intrinsic damping torque
depending on the phase difference of the coupled dynamics of YIG and Py, same as
the fieldlike torque acting along or against the Larmor
precession. The dominant mechanism for the dampinglike torque is the spin pumping from the concerted dynamics of YIG and Py \cite{TserkovnyakPRL2002,HeinrichPRL2003}

Because spin pumping is dissipative, we determine the mode with a broader (narrower) linewidth as the out-of-phase (in-phase) precession mode.
In Fig.~\ref{fig2}(a) the broader-linewidth mode exhibit a higher resonance field than the narrower-linewidth mode.
This is a signature of antiferromagnetic exchange coupling
at the YIG/Py interface \cite{KlinglerPRL2018}. From the resonance analysis we also find
that all the SiO$_2$/Py samples show lower resonance fields than
the Py samples grown on YIG \cite{supplement}, which agrees with the
antiferromagnetic nature of the YIG/Py interfacial coupling.

To reproduce the data in Fig.~\ref{fig2}(b), we introduce the
linewidths as the imaginary parts of the resonance fields in
Eq.~(\ref{eq01}):
\begin{align}\label{eq02}
\mu_0&(H_c^\pm+i\Delta H_{1/2}^\pm) = \mu_0{H_r^\text{YIG}+H_r^\text{Py} \over 2}+i\mu_0{\kappa_\text{YIG}+\kappa_\text{Py} \over 2} \nonumber \\
\pm &\sqrt{ \left(\mu_0{H_r^\text{YIG}-H_r^\text{Py} \over 2}+i\mu_0{\kappa_\text{YIG}-\kappa_\text{Py} \over 2}\right)^2 + \tilde{g}_c^2}
\end{align}
where $\kappa_\text{YIG(Py)}$ is the uncoupled linewidth of YIG (Py)
from the linear extraction (dotted lines) in Fig.~\ref{fig2}(b), and
$\tilde{g}_c=g_c+i\kappa_c$ is the complex interfacial coupling \textcolor{black}{strength}
with an additional dampinglike component $\kappa_c$ from spin pumping.

In order to show the relationship between the spin pumping from the coherent YIG-Py dynamics and the \textit{incoherent} spin pumping from the individual Py dynamics, we identify the latter as the linewidth enhancement of Py(7.5 nm), $\Delta H^{Py}_{sp}$, between the linearly extrapolated YIG/Py [red dots in Fig. 2(b)] and Si/SiO$_2$/Py [red stars in Fig. 2(b)]. {Then, we quantify the coherent dampinglike coupling strength $\kappa_c$ as $\kappa_c(\omega)=\beta\mu_0\Delta H_{sp}^\text{Py}(\omega)$, where $\beta$ is a unitless and frequency-independent value measuring the ratio between the coherent and incoherent spin pumping.}
For the best fit value, $\beta=0.82$, Eq.~(\ref{eq02}) nicely reproduces the data in
Fig.~\ref{fig2}(b). For comparison, if we set $\kappa_c(\omega)=0$ in
Eq.~(\ref{eq02}), we obtain the blue and green dashed curves, which
result in identical linewidth at $\omega_c$ as opposed to the data in
Fig.~\ref{fig2}(a).

In order to understand the physical meaning of $\tilde{g}_c$, we
consider the coupled Landau-Lifshitz-Gilbert (LLG) equations of YIG/Py
bilayer \cite{HeinrichPRL2003,TserkovnyakRMP2005,KlinglerPRL2018} in the macrospin limit:
\begin{align}\label{eq03}
&{d\mathbf{m}_i \over dt} = -\mu_0\gamma_i\mathbf{m}_i \times \mathbf{H}_{eff}+\alpha_i \mathbf{m}_i \times {d\mathbf{m}_i \over dt} \nonumber \\
-\gamma_i\mathbf{m}_i & \times {J\over M_{i}t_i}\mathbf{m}_j+\Delta \alpha_i (\mathbf{m}_i \times {d\mathbf{m}_i \over dt}-\mathbf{m}_j \times {d\mathbf{m}_j \over dt})
\end{align}
where $\mathbf{m}_{i,j}$ is the unit magnetization vector,
$\mathbf{H}_{eff}$ is the effective field including $H_B$, $H_{ex}$
and the demagnetizing field, $\alpha_i$ is the intrinsic Gilbert
damping. The index is defined as ($i$, $j$) = (1, 2) or (2, 1). In the
last two coupling terms, $J$ is the interfacial exchange energy and
$\Delta \alpha_i=\gamma_i\hbar g^{\uparrow\downarrow}/(4\pi M_{i}t_i)$
is the spin pumping damping enhancement with $g^{\uparrow\downarrow}$ the spin mixing conductance. The two terms provide the
fieldlike and dampinglike coupling torques, respectively, between
$\mathbf{m}_i$ and $\mathbf{m}_j$. To view the dampinglike coupling on
a similar footage, we define its coupling energy $J'$ as:
\begin{equation}\label{eq05}
J'(\omega) = {g^{\uparrow\downarrow} \over 4\pi}\hbar\omega
\end{equation}
Here $J'$ describes the number of quantum channels per unit area
($g^{\uparrow\downarrow}$) for magnons ($\hbar\omega$) to pass through
\cite{TserkovnyakPRL2002,TserkovnyakRMP2005}; similarly, $J$ describes
the number and strength of exchange bonds between YIG and Py per unit area. From
the definition, \textcolor{black}{we can express the spin pumping linewidth enhancement as
$\mu_0\Delta H_{sp}^i(\omega)=J'(\omega)/M_it_i$}, in
pair with the exchange field term in Eq.~(\ref{eq03}). By solving
Eq.~(\ref{eq03}) we find:
\begin{subequations}\label{eq04}
\begin{align}
&\kappa_i(\omega) = {\alpha_i\omega \over \gamma_i} + \color{black}{J'(\omega)\over M_it_i}\\
g_c =& f(\omega_c)\cdot\sqrt{{J\over M_1t_1}\cdot {J\over M_2t_2}} \\
\kappa_c(\omega_c) =& f(\omega_c)\cdot\sqrt{{J'(\omega_c)\over M_1t_1}\cdot {J'(\omega_c)\over M_2t_2}}
\end{align}
\end{subequations}
with the dimensionless factor $f(\omega)$ accounting for the precession elliptical asymmetry.
$f(\omega)=1$ for identical ellipticity ($M_1=M_2$) and $f(\omega_c)=0.9$ in the case of YIG and Py;
See the Supplmental Information for details \cite{supplement}.

\begin{figure}[htb]
 \centering
 \includegraphics[width=3.2in]{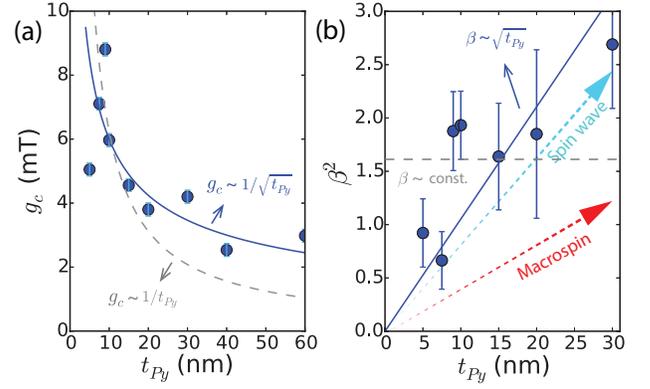}
 \caption{(a) Extracted $g_c$ as a function of $t_{Py}$. (b) Extracted $\beta^2$ as a function
   of $t_{Py}$. {In both figures, the solid and dashed curves are the fits of data to the coherent and incoherent models, respectively. In (b), the red and cyan dotted arrows show the theoretical predictions for the coherent models based on the macrospin and spin wave approximations, respectively.} Error bars indicate single standard deviations found from the fits to the lineshape.}
 \label{fig3}
\end{figure}

Eq.~(\ref{eq04}) shows that both $g_c$ and $\kappa_c$($\omega_c$) are proportional to $1/\sqrt{t_i}$,
which comes from the geometric averaging of the coupled magnetization
dynamics. This is in contrast to the $1/t_i$ dependence of the
uncoupled exchange field and spin pumping damping enhancement for a
single layer, as shown in Eq.~(\ref{eq04}a). In Fig.~\ref{fig3}(a), a good fitting of $g_c$ to $1/\sqrt{t_\text{Py}}$ rather than $1/t_\text{Py}$ validates the model. In the limit of zero Py thickness, the model breaks down due to the significance of boundary pinning and the assumption of macrospin dynamics, {as reflected in the reduction of $g_c$ at $t_\text{Py}=5$ nm.}

For the dampinglike coupling, {we plot $\beta$ instead as a function of $t_\text{Py}$ in order to minimize
the variation in the quality of interfacial coupling and
the frequency dependence of $\kappa_c$($\omega_c$)}. By taking the ratio between $\kappa_c$($\omega_c$) and $\mu_0\Delta H_{sp}^\text{Py}$($\omega_c$) from the analytical model, we obtain the macrospin expression
$\beta = f(\omega_c)\sqrt{M_\text{Py}t_\text{Py}/M_\text{YIG}t_\text{YIG}}$ with $f(\omega_c)=0.9$.
Fig.~\ref{fig3}(b) shows that the extracted $\beta^2$ varies linearly with $t_\text{Py}$,
{rather than being independent of it as would be expected for incoherent spin pumping}. {The fit is not perfect, which may be caused by i) the variation of inhomogeneous broadening of Py in YIG/Py bilayers, or ii) the multi-peak lineshapes in YIG (see YIG $n=0$ lineshapes in Fig. \ref{fig1}b) due to possible damage during the ion milling process.}

If we calculate $\beta$ from the macrospin approximation, the prediction, shown in the red dashed arrow in Fig.~\ref{fig3}(b), differs significantly from the experimental data. To
account for the difference, we consider a spin wave model
for the YIG/Py bilayer, where finite wavenumbers exist in both layers
and are determined from the boundary condition \cite{HillebrandsPRB1990}. For simplicity, we
consider free pinning at the two exterior surfaces of YIG and Py and
Hoffmann exchange boundary conditions for the interior interface of
YIG/Py \cite{HoffmannJAP1970}. From the spin wave model, we find an
additional factor of $\sqrt{2}$ in Eqs. (\ref{eq04}b)
and (\ref{eq04}c); see the Supplemental Information for details
\cite{supplement}. This factor arises because the nonuniform profile of the PSSW
mode in YIG reduces the effective mode volume by a factor of two
compared with the uniform mode. A similar effect has been previously
discussed in spin pumping from PSSW modes
\cite{KapelrudPRL2013,LiPRL2016}. In Fig.~\ref{fig3}(b) the
theoretical calculation from the spin wave model (cyan dashed arrow) {is close}
to the experimental values. {This is an additional evidence of the coherent spin pumping in YIG/Py bilayers.}

%quantitative matches to the experiments. Therefore, we conclude that the dampinglike
%coupling in YIG/Py bilayers are dominated by the mutual spin pumping from spin wave modes.

\begin{figure}[htb]
 \centering
 \includegraphics[width=3.0in]{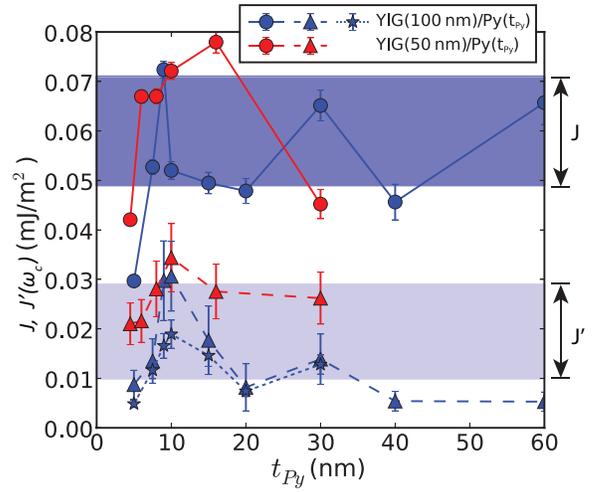}
 \caption{ Thickness dependence of $J$ (circles) and $J'$ (triangles),
   which are calculated from $g_c$ and $\kappa_c(\omega_c)$, respectively. Blue
   points denote the results for YIG(100~nm)/Py($t_{Py}$) and red points for
   YIG(50~nm)/Py($t_{Py}$). The blue stars denote $J'_{sp}$, in which $\Delta H^{Py}_{sp}(\omega_c)$
   is calculated from the Py linewidth enhancement from Py($t_{Py}$) to YIG(100~nm)/Py($t_{Py}$).
   Error bars indicate single standard deviations found from the fits to the lineshape.}
 \label{fig4}
\end{figure}

Fig.~\ref{fig4} compares the values of $J$ and $J'$ obtained from the
hybrid dynamics. For convenience we estimate the value of $J'$ from Eq.~(\ref{eq04}c), as
$J'(\omega_c) = \kappa_c(\omega_c)/f(\omega_c)\cdot
\sqrt{M_\text{YIG}t_\text{YIG}M_\text{Py}t_\text{Py}/2}$. 
Noting the frequency dependence of $J’(\omega)$, all the values of $J’(\omega)$ in this work are obtained around $\omega_c/2\pi=9$ GHz.
We can also calculate $J'(\omega_c)$ from uncoupled spin
pumping effect, as
$J'_{sp}(\omega_c)=\mu_0\Delta H_{sp}^\text{Py}(\omega_c)\cdot
M_\text{Py}t_\text{Py}$. \textcolor{black}{For the YIG/Py interface, the
value of $J$ stays at the same level; the value of $J'(\omega_c)$ fluctuates
with samples but is well aligned with $J'_{sp}(\omega_c)$, which again
supports that the dampinglike interfacial coupling comes from spin
pumping.} Furthermore, we have also repeated the experiments for
a thinner YIG(50~nm)/Py(t) sample series and obtained similar values of
$J$ and $J'(\omega_c)$, as shown in Fig. \ref{fig4}.

Table \ref{table1} summarizes the values of $J$, $J'$ and
$g^{\uparrow\downarrow}$ for YIG/Py interface, 
where $J’$ is taken from the vicinity of $\omega_c/2\pi=9$ GHz and
$g^{\uparrow\downarrow}$ is calculated from $J'(\omega_c)$ by Eq.~(\ref{eq05}). 
The value of $J$ is much smaller than a perfect exchange coupled
interface, {which is not surprising given the complicated and uncharacterized nature of the YIG/Py interface.}
{For Py, the interfacial exchange energy can be estimated \cite{HillebrandsPRB1990} by $2A_{ex}/a$, where for Py $A_{ex}=12$ pJ/m \cite{LiPRL2016} and the lattice parameter $a=0.36$~nm.} We find $2A_{ex}/a=68$~mJ/m$^2$, three
orders of magnitude larger than $J$. Comparing with similar
interfaces, our reported $J$ is similar to YIG/Ni (0.03~mJ/m$^2$
\cite{ChenPRL2018}) and smaller than YIG/Co (0.4~mJ/m$^2$
\cite{KlinglerPRL2018}). A different interlayer exchange
coupling from Ruderman-Kittel-Kasuya-Yosida interaction may generate a
larger $J$ \cite{ParkinPRL1990,BelmeguenaiPRB2007,FallarinoAPL2016}
but a smaller $g^{\uparrow\downarrow}$ \cite{YangAPL2016}. There could
also be a fieldlike contribution of $J$ from $g^{\uparrow\downarrow}$
\cite{WeiZhangPRB2015,SklenarPRB2015,NanPRB2015,KlinglerPRL2018,BuhrmanPRL2019}. But since the
exchange $J$ dominates in the coupled dynamics, it is difficult to
distinguish the spin mixing conductance contribution in our
experiments.

\begin{table}[ht]
\centering
%\begin{tabular}{cccccc}
\begin{tabular}{>{\centering\arraybackslash}m{0.8in} >{\centering\arraybackslash}m{0.8in} >{\centering\arraybackslash}m{0.8in} >{\centering\arraybackslash}m{0.8in} }
\hline
\hline
& $J$(mJ/m$^2$) & $J'$(mJ/m$^2$) & $g^{\uparrow\downarrow}$ (nm$^{-2}$)  \\
\hline
YIG/Py&  $0.060\pm0.011$ & $0.019\pm0.009$ & $42\pm21$   \\
\hline
\end{tabular}
\caption{Fieldlike, dampinglike coupling energy and spin mixing conductance for the YIG/Py interface. The value of $J’$ is calculated around $\omega_c/2\pi=9$ GHz}
\label{table1}
\end{table}

%{In Fig. \ref{fig4} both $J$ and $J'$ start to decrease as the Py thickness becomes smaller. For the exchange coupling $J$, the reduction at $t_{Py}=5$ nm indicate the expiration of interfacial exchange coupling, which may be linked to the nonuniform static magnetization profile of Py. For the dampinglike coupling, the decrease of $J'$ from $t_{Py}=10$ nm suggests the characterization length of spin current penetration, below which the effective spin mixing conductance start to decrease \cite{TserkovnyakRMP2005}.}

%We note that the value of $\Delta H_{sp}^\text{Py}$ is not proportional to $\omega$. This indicates that the dampinglike coupling strength $\kappa_c$ cannot be understood fully by a macrospin model and might involve some extrinsic broadening mechanism, such as inhomogeneous broadening or two-magnon scattering. Nevertheless, the good match between $J'$ and $J'_{sp}$ suggests the coherence of the dampinglike coupling in the coupling dynamics, which supports the role of spin pumping rather than the other two incoherent broadening mechanisms.

In conclusion, we have characterized the dampinglike coupling torque
between two exchange-coupled ferromagnetic thin films. By exciting the
hybrid dynamics in the strong coupling regime, this dampinglike torque
can either increase or suppress the total damping in the out-of-phase
or in-phase mode, respectively. The origin of the dampinglike torque
is the coherent spin pumping from the coupling magnetization
dynamics. Our results reveal new insight for tuning the coherence in
magnon-magnon hybrid dynamics and are important for magnon-based
coherent information processing.

Work at Argonne on sample preparation was supported by the U.S. DOE, Office of Science, Office of Basic Energy Sciences, Materials Science and Engineering Division under Contract No. DE-AC02-06CH11357, while work at Argonne and National Institute of Standards and Technology (NIST) on data analysis and theoretical modeling was supported as part of Quantum Materials for Energy Efficient Neuromorphic Computing, an Energy Frontier Research Center funded by the U.S. DOE, Office of Science. Work on experimental design at Oakland University was supported by AFOSR under grant no. FA9550-19-1-0254 and the NIST Center for Nanoscale Science and Technology, Award No. 70NANB14H209, through the University of Maryland. Work on microwave spectroscopy at Columbia University was supported by NSF under grant NSF-DMR1411160.

%\bibliography{ref,ref_QIS,ref_YIG_magnon}


\end{document}